\documentclass[a4paper,11pt]{article}
\usepackage{pos}
\usepackage{graphicx}
\usepackage{dcolumn}
\usepackage{bm}
\usepackage{amsmath}
\usepackage{braket}
\usepackage{slashed}
\usepackage{epstopdf}
\usepackage{placeins}
\usepackage{multirow}
\usepackage{makecell}
\usepackage{soul}
\usepackage{braket}
\usepackage[shortlabels]{enumitem}
\usepackage{placeins}

\usepackage[font=small, labelfont=bf, textfont={small,it}]{caption}

\newcommand{\beq}{\begin{eqnarray}}
	\newcommand{\eeq}{\end{eqnarray}}
\newcommand{\beqnn}{\begin{eqnarray*}}
	\newcommand{\eeqnn}{\end{eqnarray*}}

\newcommand{\Tr}{\ensuremath{\mathrm{Tr}}}

\newcommand{\cool}{\ensuremath{\mathrm{cool}}}

\newcommand{\SU}{\ensuremath{\mathrm{SU}}}

\newcommand{\sphal}{\mathrm{Sphal}}

\title{An update on the determination of the sphaleron rate in finite temperature QCD}

\author[a,b]{Nicolò Bellini}
\author[c]{Claudio Bonanno}
\author*[a,b,d]{Francesco D'Angelo}
\author[a,b]{Massimo D'Elia}
\author[a,b]{Andrea Giorgieri}
\author[e,f]{Lorenzo Maio}

\affiliation[a]{Università di Pisa, Largo B.~Pontecorvo 3, I-56127 Pisa, Italy}

\affiliation[b]{INFN, Sezione di Pisa, Largo B.~Pontecorvo 3, I-56127 Pisa, Italy}

\affiliation[c]{Instituto de F\'isica Te\'orica UAM-CSIC, c/ Nicol\'as Cabrera 13-15, Universidad Aut\'onoma de Madrid, Cantoblanco, E-28049 Madrid, Spain}

\affiliation[d]{INFN, Sezione di Roma Tre, Via della Vasca Navale 84, I-00146 Rome, Italy}

\affiliation[e]{Aix Marseille Univ., Université de Toulon, CNRS, CPT, Marseille 13009, France}

\affiliation[f]{Dipartimento di Fisica, Università di Roma “Tor Vergata” and INFN, Sezione di Roma Tor Vergata, Via della Ricerca Scientifica 1, I-00133 Rome, Italy}

\emailAdd{francesco.dangelo@phd.unipi.it}

\abstract{The sphaleron rate is a key phenomenological quantity both for the axion thermal production in the Early Universe and the Chiral Magnetic Effect occurring in the Quark-Gluon Plasma in presence of a background magnetic field. In this talk we present an extension of our recent determination of the sphaleron rate, in the $\SU(3)$ gauge theory, based on the determination of the two-point function of the topological charge density at finite temperature.}

\FullConference{The 41st International Symposium on Lattice Field Theory (LATTICE2024)\\
 28 July - 3 August 2024\\
Liverpool, UK\\}

\begin{document}
\maketitle

\section{Introduction}\label{sec:introduction}

The study of real time QCD topological transitions at finite temperature, described by the \textit{strong sphalerons}, is extremely important not only for a better understanding of the QCD vacuum properties, but also for their phenomenological implications. The relevant physical quantity that describes these thermal processes is the \textit{strong sphaleron rate}, defined as
\beq\label{eq:sphal_rate_def}
	\Gamma_\sphal = \underset{t_{\mathrm{M}}\to\infty}{\underset{V_s\to\infty}{\lim}} \, \frac{1}{V_s t_{\mathrm{M}}}\left\langle\left[\int_0^{t_{\mathrm{M}}} d t_{\mathrm{M}}' \int_{V_s} d^3x \, q(t_{\mathrm{M}}', \vec{x})\right]^2\right\rangle=\int d t_{\mathrm{M}} d^3x \braket{q(t_{\mathrm{M}},\vec{x}) q(0,\vec{0})},
\eeq
where $t_{\mathrm{M}}$ is the Minkowskian time and 
\beq\label{eq:top_charge_dens}
	q(x)=\frac{1}{32\pi^2}\epsilon_{\mu\nu\rho\sigma}\operatorname{Tr}\{G^{\mu\nu}(x)G^{\rho\sigma}(x)\}
\eeq
is the topological charge density operator, with $G_{\mu\nu}$ being the gauge field strength.

The strong sphaleron rate plays an important phenomenological role in different contexts such as Quark-Gluon Plasma (QGP) and axion physics. For instance, a non vanishing $\Gamma_\sphal$ is related to local imbalances in the number of left/right-handed quark species in the QGP, giving rise to the well-known Chiral Magnetic Effect~\cite{Fukushima:2008xe,Kharzeev:2013ffa,Astrakhantsev:2019zkr,Almirante:2024lqn}, i.e., the appearance of an electric current in the same direction of a background magnetic field. On the other side, as pointed out in a recent paper~\cite{Notari:2022ffe}, this quantity is also related to the thermal production of axions in the Early Universe, as it enters the Boltzmann equation for the 3-momentum dependent axion distribution function.

Since sphaleron dynamics is strictly non-perturbative, the lattice regularization provides a useful framework for the determination of $\Gamma_\sphal$. However, Monte Carlo simulations on the lattice require an Euclidean formulation of the theory, so the real time definition in Eq.~\ref{eq:sphal_rate_def} can not be directly employed. Different strategies can be adopted, and most of them are based on the fact that $\Gamma_\sphal$ is related to the \textit{spectral density} $\rho(\omega)$ of the Euclidean topological charge density time correlator $G(t)$ via the \textit{Kubo formula} ($T$ is the temperature)
\beq\label{eq:kubo_formula}
	\Gamma_\sphal = 2T\lim_{\omega\to0}\frac{\rho(\omega)}{\omega}.
\eeq
More precisely, the quantity directly accessible on the lattice is the Euclidean time correlator of the topological charge density operator $G(t)$, related to the spectral density $\rho(\omega)$ via the integral relation (now $t$ is the imaginary time)
\beq\label{eq:lat_corr_int_relation}
	G(t)\equiv\int d^3x \langle q(t,\vec{x})q(0,\vec{0})\rangle = -\int_0^\infty \frac{d\omega}{\pi}\rho(\omega) \frac{\cosh\left[\frac{\omega}{2T}-\omega t\right]}{\sinh\left[\frac{\omega}{2T}\right]}.
\eeq
The strategy is now clear: firstly, one can determine $G(t)$ from lattice simulations and then invert Eq.~\ref{eq:lat_corr_int_relation} to obtain $\Gamma_\sphal$. However, this kind of inversion is a very complicated task, being a mathematical ill-posed problem\footnote{We refer the reader to Refs.~\cite{Rothkopf:2022fyo,Aarts:2023vsf} for some recent reviews on the topic.}.

Different approaches have been developed to attack this kind of problem~\cite{Boito:2022njs,Horak:2021syv,DelDebbio:2021whr,Candido:2023nnb,Tikhonov:1963aaa,Astrakhantsev:2018oue,Astrakhantsev:2019zkr,BackusGilbert1968:aaa,Brandt:2015aqk,Brandt:2015sxa,Hansen:2019idp,ExtendedTwistedMassCollaborationETMC:2022sta,Frezzotti:2023nun,Evangelista:2023fmt}. Among them, we adopt the well-known Hansen--Lupo--Tantalo (HLT)~\cite{Hansen:2019idp} modification of the Backus--Gilbert method~\cite{BackusGilbert1968:aaa}, which allows to estimate the $g_t(0)$ coefficients and finally write the spectral density as a linear combination of the correlator at different times
\beq\label{eq:sphal_rate_from_hlt}
	\frac{\Gamma_\sphal}{2T}=\left[\frac{\bar{\rho}(\bar{\omega})}{\bar{\omega}}\right]_{\bar{\omega}=0}=-\pi\sum_{t=0}^{1/T}g_t(0)G(t).
\eeq

\begin{figure}
	\centering
	\includegraphics[width=0.45\textwidth]{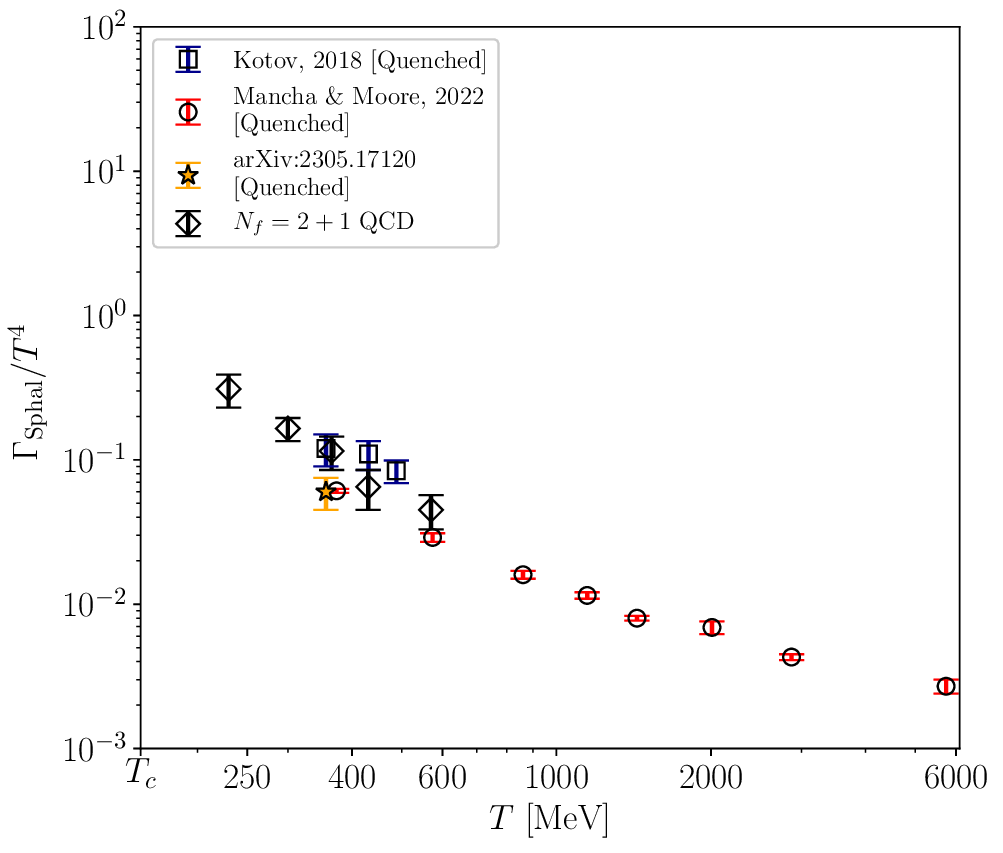}
	\includegraphics[width=0.45\textwidth]{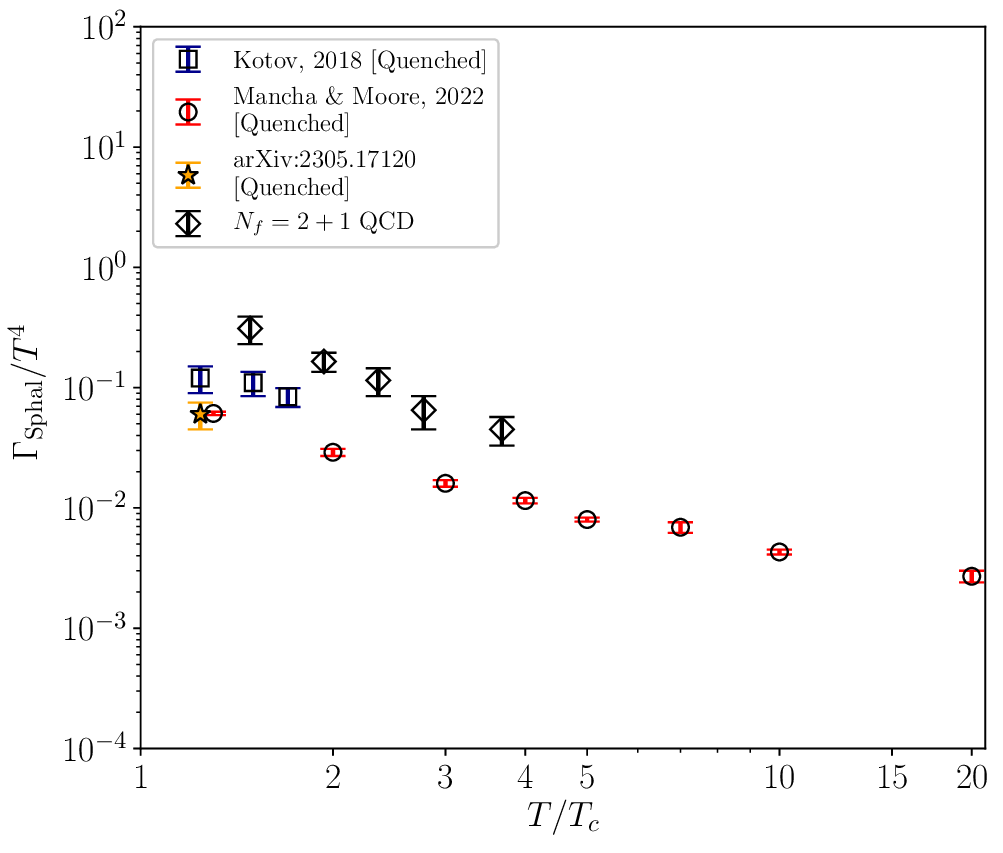}
	\caption{$N_f=2+1$ QCD sphaleron rate behaviour, represented with diamond points taken from Ref.~\cite{Bonanno:2023thi}, as a function of the temperature. A comparison with some previous quenched results is also done: square points are taken from Refs.~\cite{Kotov:2018aaa,Kotov:2019bt}, round markers from Ref.~\cite{BarrosoMancha:2022mbj} and, finally, the starred one from Ref.~\cite{Bonanno:2023ljc}. Left: x-axis expressed in terms of the absolute temperature in $\mathrm{MeV}$. Right: x-axis expressed in terms of $T/T_c$, with $T_c=155~\mathrm{MeV}$ and $T_c=287~\mathrm{MeV}$ for full QCD and the quenched theory, respectively.}
	\label{fig:full_qcd_results}
\end{figure}

In the literature only few attempts for the computation of $\Gamma_\sphal$ are reported, and most of them regard the quenched theory~\cite{Kotov:2018aaa,Kotov:2019bt,BarrosoMancha:2022mbj}. The QCD case, the relevant one for the axion phenomenology, has recently been investigated for the first time~\cite{Bonanno:2023thi} for 5 different temperatures in the range $200~\mathrm{MeV}\lesssim T \lesssim 600~\mathrm{MeV}$ by adopting a methodology already tested in the quenched case~\cite{Bonanno:2023ljc}, which turned out to be consistent with the previous determinations in the literature. A summary of the QCD results, and a comparison with the quenched determinations, can be found in Fig.~\ref{fig:full_qcd_results}. 

In this talk, we focus on the extension of the sphaleron rate computation to the non-zero spatial momentum case that is relevant for axion phenomenology. First of all, following Ref.~\cite{Notari:2022ffe}, we motivate this kind of study. Then, after discussing the numerical setup, we show some very preliminary results for the quenched theory. Finally, we conclude with some comments and future outlooks.

\section{Axion rate from strong sphalerons}

The axion distribution function $f_{\vec{p}}$ with comoving 3-momentum $\vec{p}$ in the QCD thermal bath can be obtained by solving the Boltzmann equation
\beq\label{eq:boltzmann}
	\frac{df_{\vec{p}}}{dt}=(1+f_{\vec{p}})\Gamma^{<}-f_{\vec{p}}\Gamma^{>}\ ,
\eeq
where the axion is assumed to have a negligible mass (so $p^\mu=(E=\vert\vec{p}\vert,\vec{p})$), while $\Gamma^<$ and $\Gamma^>$ describe the energy dependent creation/destruction rates. At the thermal equilibrium in the QCD bath, the following non-perturbative relation holds
\beq
\Gamma^> = \exp\left(\frac{E}{T}\right)\Gamma^< =\frac{\Gamma^{>}_{\mathrm{top}}}{2Ef_a^2}\ ,
\eeq
where $T$ is the temperature and $\Gamma^{>}_{\mathrm{top}}$ is the \textit{topological rate} related to the $2$-point correlation function of the topological charge density operator $q(x)$
\beq
\Gamma^{>}_{\mathrm{top}}(p^\mu)\equiv\int d^4x\ e^{ip^\mu x_\mu}\langle q(x^\mu)q(0)\rangle\ .
\eeq
It is worth noticing that $\Gamma^{>}_{\mathrm{top}}(p^\mu=0)=\Gamma_\sphal$ (compare with Eq.~\ref{eq:sphal_rate_def}).

It is now clear the motivation of the extension of the sphaleron rate computation to the non-zero momentum case: what is really needed to put more stringent bounds on the axion mass is the 3-momentum dependence of $\Gamma^>_\mathrm{top}(\vec{p})$, i.e., the \textit{total} axion production rate. Moreover, the authors of Ref.~\cite{Notari:2022ffe} also argue that for $T\lesssim5~\mathrm{GeV}$ the contribution to $\Gamma^>_\mathrm{top}(\vec{p})$ is dominated by sphalerons. This is the reason why the topological rate is expected to be constant for values of 3-momenta $\vert \vec{p} \vert$ smaller that the 3-momentum  associated to the sphaleron size $\vert \vec{p}_s\vert\sim N_c\alpha_s T$
\beq
	&\Gamma^>_\mathrm{top}(E=\vert\vec{p}\vert\lesssim\vert\vec{p}_s\vert)\simeq\Gamma^>_\mathrm{top}(E=0)\equiv\Gamma_\sphal,
\eeq 
while it is expected to sharply decay for $\vert \vec{p} \vert \gtrsim \vert \vec{p}_s\vert$.
 
However, in order to determine explicitly the dependence of the topological rate on the 3-momentum, a very useful tool is provided by lattice QCD simulations. The approach is quite similar to the one sketched in Sec.~\ref{sec:introduction}. On the lattice, one can access the spatial Fourier transform of the Euclidean time-correlator of the topological charge density operator $G^{\vec{p}}(t)$, that can be expressed in terms of the spectral density $\rho(\omega,\vec{p})$ via the following integral relation
\beq
	G^{\vec{p}}(t)\equiv\int d^3x\ e^{i\vec{p}\cdot\vec{x}}\langle q(t,\vec{x})q(0,\vec{0})\rangle = -\int \frac{d\omega}{\pi}\rho(\omega,\vec{p})\frac{\cosh\left[\frac{\omega}{2T}-\omega t\right]}{\sinh\left[\frac{\omega}{2T}\right]}.
\eeq
Once $G^{\vec{p}}(t)$ has been computed on the lattice, one can rely on the Kubo formula for the non-zero momentum case, invert the correlator via the HLT Backus--Gilbert and obtain the non-zero momentum topological rate~\cite{Son:2002sd}
\beq\label{eq:non_zero_mom_topological_rate}
	\Gamma^{>}_{\mathrm{top}}(\vert\vec{p}\vert)=\left[\coth\left(\frac{\bar{\omega}}{2T}\right)\bar{\rho}(\bar{\omega},\vec{p})\right]_{\bar{\omega}=\vert\vec{p}\vert} = \coth\left(\frac{\vert\vec{p}\vert}{2T}\right)\left[-\pi \sum^{1/T}_{t=0}g_t(\bar{\omega}=\vert\vec{p}\vert)G^{\vec{p}}(t)\right],
\eeq
where the inversion now has to be performed in $\omega=\vert\vec{p}\vert$ because of the axion dispersion relation in the massless approximation. It is worth noticing that Eq.~\ref{eq:non_zero_mom_topological_rate} reduces to Eq.~\ref{eq:kubo_formula} in the limit $\omega\to0$, being $\coth(x)\sim1/x$ for $x\to0$.

\section{Numerical results}

In the following, we will show some preliminary results about the 3-momentum dependence of $G^{\vec{p}}(t)$ in the quenched case at $T\simeq1.24~T_c$ (the same of Refs.~\cite{Kotov:2018aaa,Bonanno:2023ljc}).

We perform Monte Carlo simulations of the $\SU(3)$ gauge theory at $T\simeq1.24~T_c$, with $T_c\simeq287~\mathrm{MeV}$. We adopt the same lattice setup of Ref.~\cite{Bonanno:2023ljc}: the gauge sector is discretized with the standard Wilson action on $N_s^3\times N_t$ lattices, and each Monte Carlo step consists of 1 lattice sweep of Over-Heat-Bath~\cite{Creutz:1980zw,Kennedy:1985nu} followed by 4 sweeps of Over-Relaxation~\cite{Creutz:1987xi}, both implemented \textit{à la} Cabibbo--Marinari~\cite{Cabibbo:1982zn}. The lattice parameters, summarized in Tab.~\ref{tab:simulation_summary_sphal_rate_non_zero_mom}, are the same of Ref.~\cite{Bonanno:2023ljc}, with the exception that for this study we adopt a larger aspect ratio $N_s/N_t=4$, to have a finer spacing among discretized lattice momenta.
\begin{table}[!t]
	\begin{center}
		\begin{tabular}{|c|c|c|c|c|c|}
			\hline
			$N_s$ & $N_t$ & $\beta$ & $a~\mathrm{[fm]}$ & $L~\mathrm{[fm]}$ & $T/T_c$ \\
			\hline
			$56$ & $14$ & $6.559$ & $0.03948(58)$ & $2.211(32)$ & $1.242(18)$ \\
			$64$ & $16$ & $6.665$ & $0.03450(50)$ & $2.208(32)$ & $1.244(18)$ \\
			$80$ & $20$ & $6.836$ & $0.02759(40)$ & $2.207(32)$ & $1.244(18)$ \\
			\hline
		\end{tabular}
	\end{center}
	\caption{Summary of simulation parameters for the quenched case. For more details about the scale setting, we refer the reader to Ref.~\cite{Bonanno:2023ljc}. Scale setting obtained by interpolating results for $a/r_0$ of Ref.~\cite{Necco:2001xg}, where $r_0 = 0.472(5)~\mathrm{fm}$~\cite{Sommer:2014mea} is the Sommer scale.}
	\label{tab:simulation_summary_sphal_rate_non_zero_mom}
\end{table}

The topological charge density operator is discretized with the standard clover definition, which has a definite parity
\beq\label{eq:clover_definition}
	q_L(n_t;\vec{n})=\frac{-1}{2^9\pi^2}\sum_{\mu\nu\rho\sigma=\pm1}^{\pm4}\varepsilon_{\mu\nu\rho\sigma}\Tr\{\Pi_{\mu\nu}(n_t; \vec{n})\Pi_{\rho\sigma}(n_t;\vec{n})\},
\eeq
where $\Pi_{\mu\nu}(n_t; \vec{n})$ is the plaquette starting from $n\equiv(n_t,\vec{n})$ and extending in the $\mu-\nu$ plane, and $\varepsilon_{(-\mu)\nu\rho\sigma}=-\varepsilon_{\mu\nu\rho\sigma}$. In order to compute $G^{\vec{p}}(t)$ on the lattice, it is useful to define the 3-momentum dependent time profile $Q^{\vec{p}}_L(n_t)$
\beq\label{eq:mom_dep_time_profile}
	Q_L^{\vec{p}}(n_t)\equiv\sum_{\vec{n}}e^{i\vec{p}\cdot\vec{n}}q_L(n_t;\vec{n}),\qquad\vec{p}=\frac{2\pi}{N_s}(k_x,k_y,k_z)\text{ with } k_i\in[0,\ldots, N_s-1],
\eeq
that allows us to easily determine the dimensionless spatial Fourier transform
\beq\label{eq:lattice_corr_def}
	\frac{G_L^{\vec{p}}(tT)}{T^5}=\frac{N_t^5}{N_s^3}\left\langle Q^{\vec{p}}_L(n_{t,1}) Q^{-\vec{p}}_L(n_{t,2}) \right\rangle,\quad tT =  \operatorname{min}\left\{\frac{\vert n_{t,1}-n_{t,2}\vert}{N_t}; 1-\frac{\vert n_{t,1}-n_{t,2}\vert}{N_t}\right\},
\eeq
where the normalized time separation $tT$ is defined taking into account the presence of periodic boundary conditions. 

Let us comment a little bit more on the choice of the external 3-momentum. Since there is not any source of anisotropy, we can choose without loss of generality a 3-momentum $\vec{p}$ oriented along a certain axis: indeed, the relevant quantity is $\vert\vec{p}\vert/T$. For our study we set $k\equiv k_x\neq0$ and $k_y=k_z=0$, i.e., $\vec{p}$ is oriented along the x-axis. However, in the future we are planning to take measurements also for momenta oriented in the y- and z-axis, in order to have a larger statistics.

The two-point function of the topological charge density operator is known to be affected by short distance singularities that give rise to an additive renormalization term~\cite{DiVecchia:1981aev,DElia:2003zne} that overcomes the physical signal in the continuum limit. In order to suppress these UV fluctuations, a smoothing procedure is required. In this study we adopt cooling~\cite{Berg:1981nw,Iwasaki:1983bv,Itoh:1984pr,Teper:1985rb,Ilgenfritz:1985dz,Campostrini:1988cy,Alles:2000sc}, characterized by a certain number $n_\cool$ of discrete steps, which can be associated to a \textit{smearing radius} $r_sT\sim \sqrt{(8/3)n_\cool}/N_t$~\cite{Bonati:2014tqa}. The correlator is finally computed by performing a continuum limit at fixed smoothing radius $r_s$ in physical units, followed by a zero smoothing limit. This is done according to a well-established procedure in the literature, and for more details we refer the reader to Refs.~\cite{Altenkort:2020axj,Bonanno:2023ljc}.

\begin{figure}[!htb]
	\centering
	\includegraphics[scale=0.40]{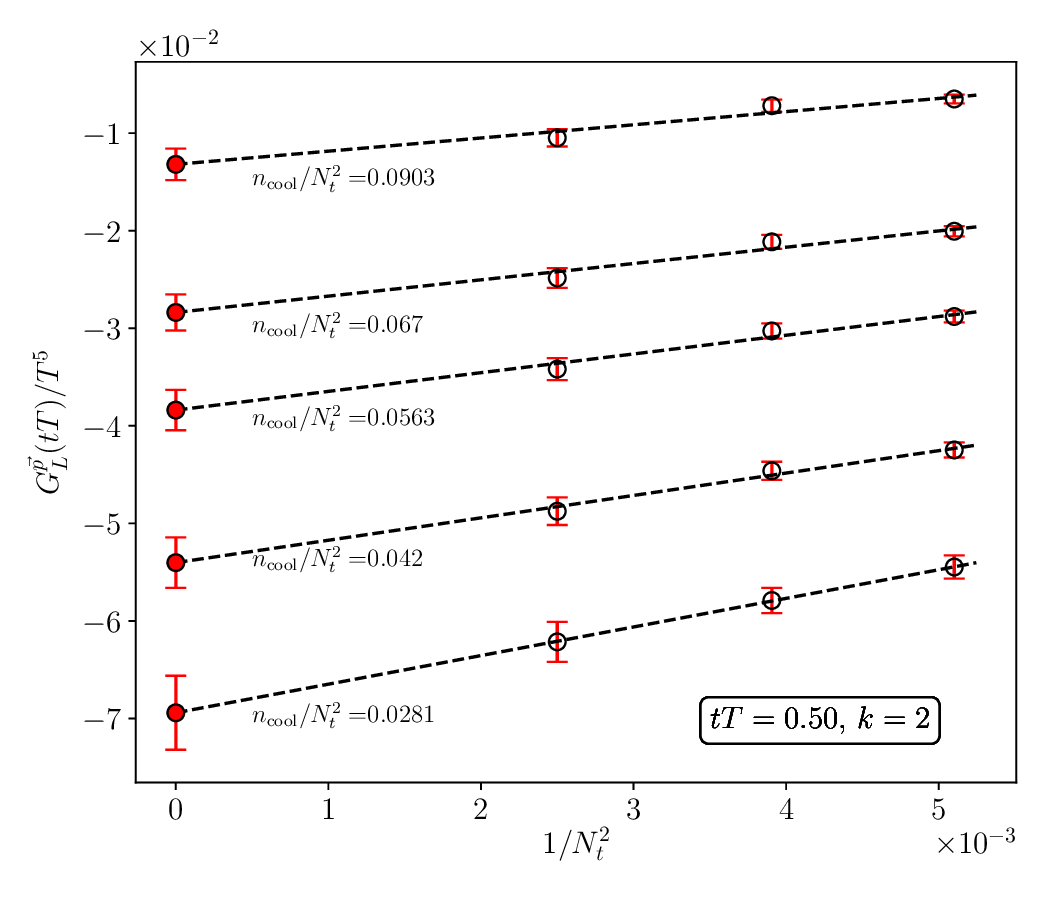}
	\includegraphics[scale=0.40]{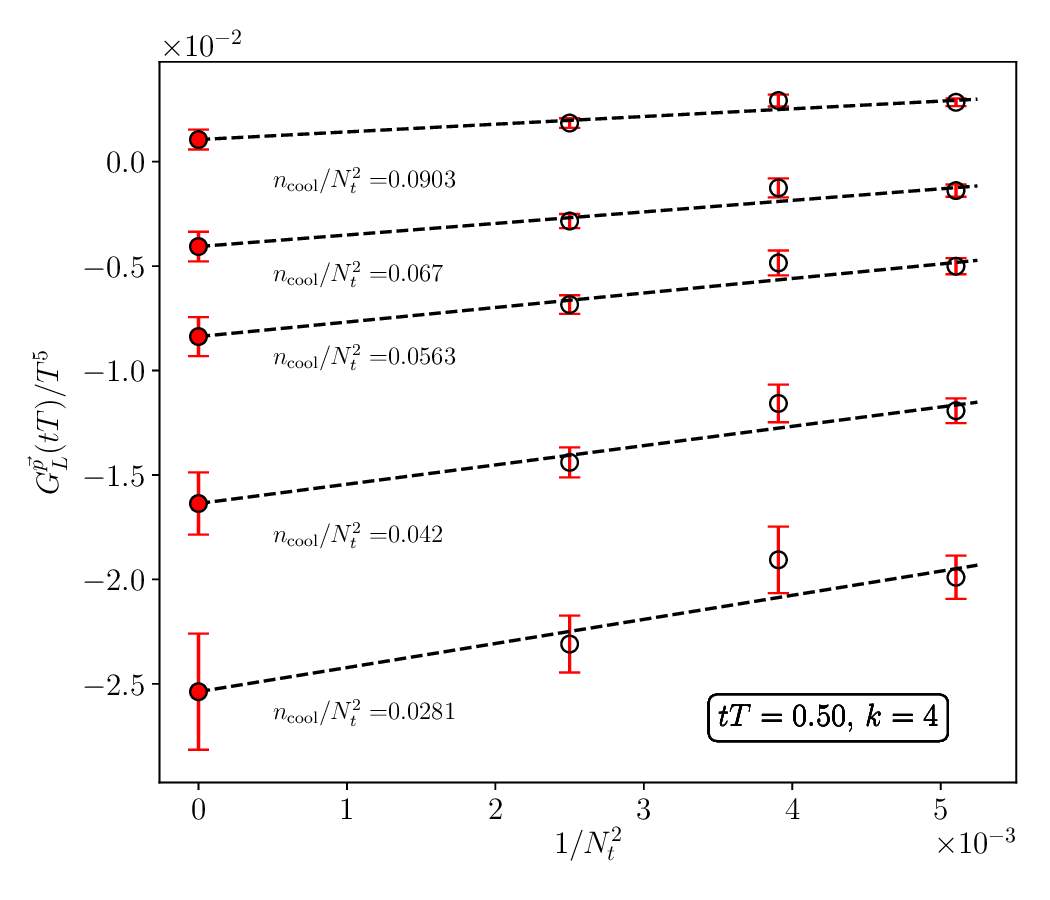}
	\caption{Continuum limit of the correlator $G_L^{\vec{p}}(tT)$, at $tT=0.5$, for some values of the smoothing radius $(r_sT)^2\propto n_\cool/N_t^2$, according to the fit function in Eq.~\ref{eq:continuum_limit_scaling}. Left: $k=2$. Right: $k=4$.}
	\label{fig:cont_lim_k_not_zero}
\end{figure}

For every value of $k$, or equivalently $\vert\vec{p}\vert/T$, the continuum limit is performed by assuming standard $O(1/N_t^2)$ corrections
\beq\label{eq:continuum_limit_scaling}
	\frac{G^{\vec{p}}_L \left(tT,N_t,\frac{n_{\mathrm{cool}}}{N_t^2}\right)}{T^5}=\frac{G^{\vec{p}}\left(tT,\frac{n_{\mathrm{cool}}}{N_t^2}\right)}{T^5}+b^{\vec{p}}\left(tT,\frac{n_{\mathrm{cool}}}{N_t^2}\right)\frac{1}{N_t^2}+o\left(\frac{1}{N_t^2}\right),	
\eeq
where it has been pointed out the fact that $O(1/N_t^2)$ corrections can in principle depend on the time separation $tT$, the smoothing radius and the 3-momentum $\vec{p}$. In Fig.~\ref{fig:cont_lim_k_not_zero}, we display an example of continuum limit for $tT=0.5$ and two different values of $k$: $k=2,4$, corresponding to $\vert\vec{p}\vert/T\simeq3.14,6.28$, respectively. It is clearly visible how our data are well described by the fit function in Eq.~\ref{eq:continuum_limit_scaling}. The same conclusion still holds for other values of $k$ and $tT$.

\begin{figure}[!htb]
	\centering
	\includegraphics[scale=0.40]{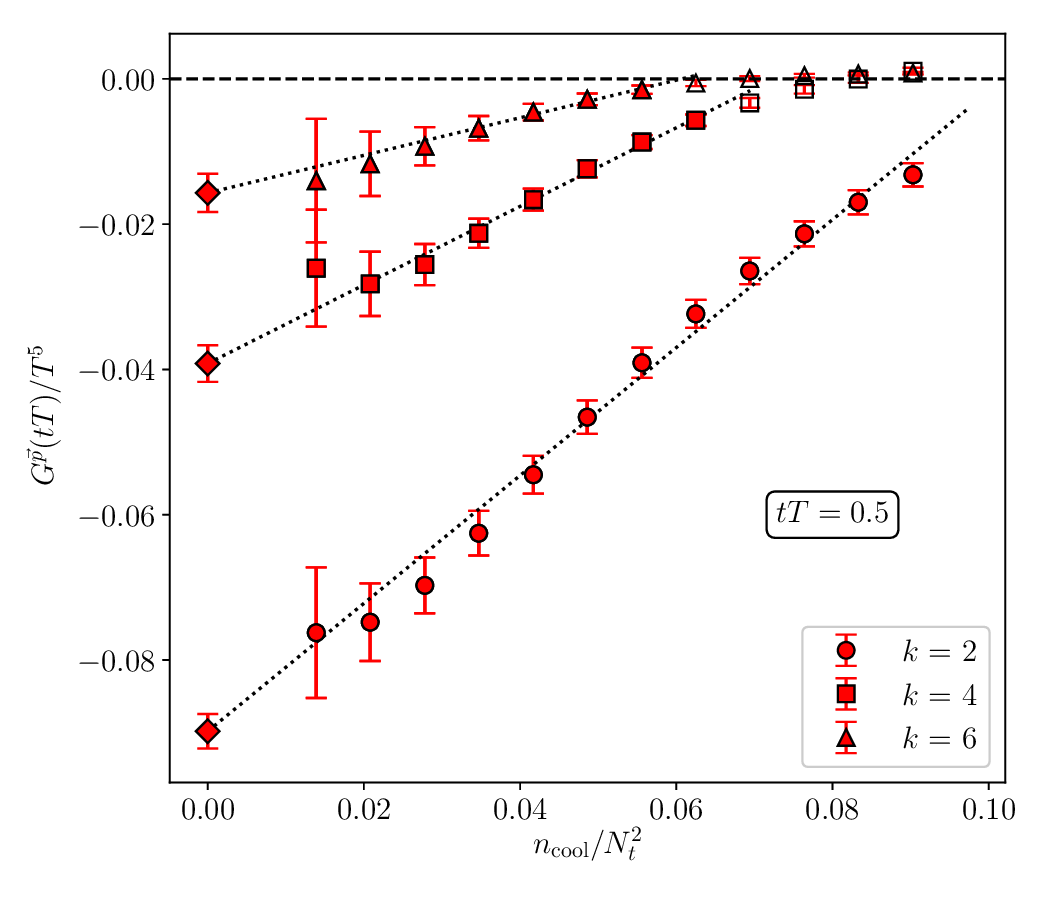}
	\includegraphics[scale=0.40]{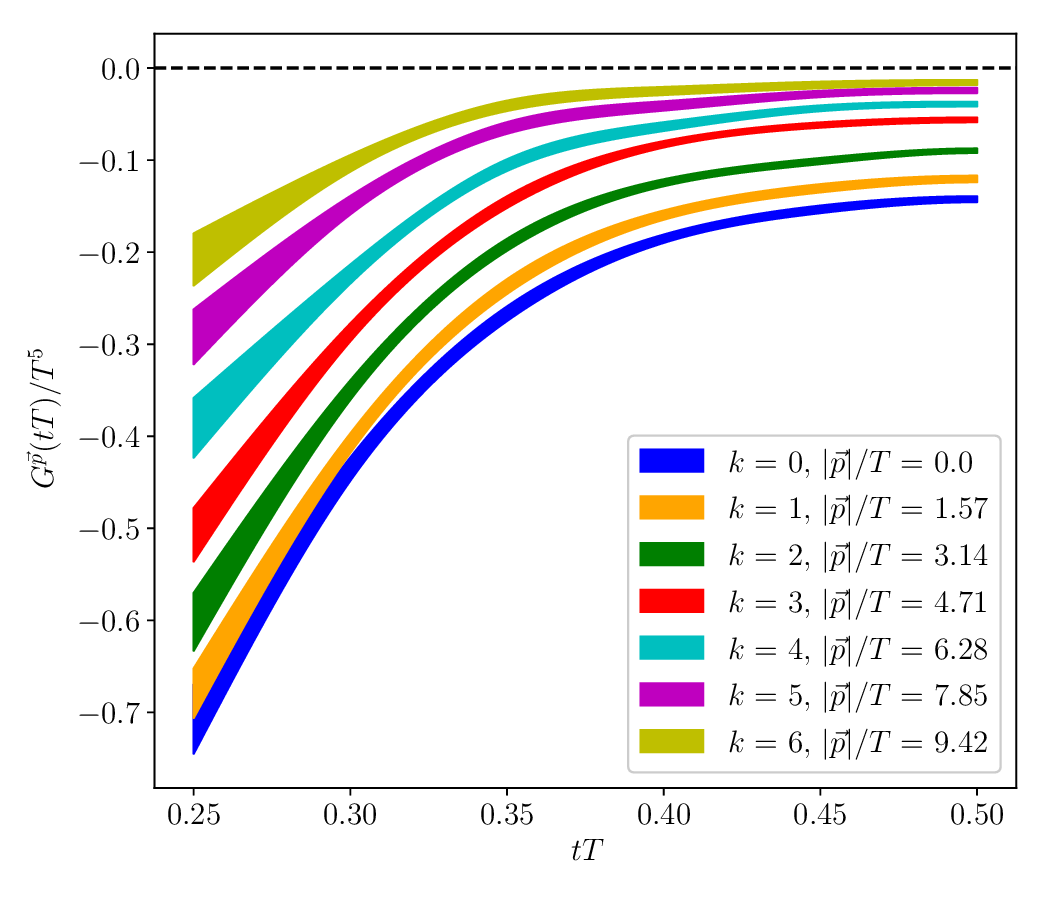}
	\caption{Left: zero smoothing limit of $G^{\vec{p}}(tT,n_\cool/N_t^2)/T^5$ at $tT=0.5$, for different values of $k$. A fit has been performed according to Eq.~\ref{eq:zero_smoothing} using only filled points. A reference line in zero is also displayed. Right: double extrapolated correlator $G^{\vec{p}}(tT)/T^5$ as a function of $\vert\vec{p}\vert/T$, with a reference line in zero displayed.}
	\label{fig:zero_smoothing_limit}
\end{figure}

Once the continuum limit has been performed, the correlator is affected by a residual dependence on the smoothing radius $r_s$, which has to be removed by performing a zero smoothing limit. According to what has been previously done in Refs.~\cite{Altenkort:2020axj,Bonanno:2023ljc}, we assume linear corrections in $n_\cool/N_t^2$.
\beq\label{eq:zero_smoothing}
	\frac{G^{\vec{p}}\left(tT,\frac{n_{\mathrm{cool}}}{N_t^2}\right)}{T^5}=\frac{G^{\vec{p}}(tT)}{T^5}+c^{\vec{p}}(tT)\frac{n_{\mathrm{cool}}}{N_t^2}.	
\eeq
When performing the $r_s\to0$ limit, one has to pay attention to the fit range adopted\footnote{We refer the reader to Refs.~\cite{Altenkort:2020axj,Bonanno:2023ljc} for detailed discussions about this issue, since in this study we adopt the same criteria.}. More precisely, one has to perform a sufficient amount of smoothing to correctly identify the topological background. This criterium allows us to determine a lower bound for the fit range: in particular, we look for a common value $n_\cool^{\mathrm{(min)}}/N_t^2$ across all ensembles, where the topological susceptibility starts exhibiting a plateau. In this study we have $n_\cool^{\mathrm{(min)}}/N_t^2\simeq0.012$, the same value of Ref.~\cite{Bonanno:2023ljc}. With respect to the upper bound $n_\cool^{\mathrm{(max)}}/N_t^2$, we adopt the requirement for $r_s$ not to be larger than the time separation $tT$, in order not to have overlapping sources in the correlator that would lead to unphysical results. As it can be appreciated from le left hand side plot of Fig.~\ref{fig:zero_smoothing_limit}, where some examples of zero smoothing extrapolations are shown, Eq.~\ref{eq:zero_smoothing} well describes our data. However, it is worth noticing that, when $k$ becomes larger, the upper bound in the fit range gets smaller, i.e., deviations from linearity occur for smaller values of $n_\cool$.

In the right hand side plot of Fig.~\ref{fig:zero_smoothing_limit}, the final result, i.e., the double extrapolated correlator $G^{\vec{p}}(tT)/T^5$, is shown for different values of $\vert\vec{p}\vert/T$. First of all, the double extrapolated correlator is negative for every $tT>0$, as it should be according to the reflection positivity property~\cite{Vicari:1999xx}. As $\vert\vec{p}\vert/T$ is increased, the correlator is suppressed, signalling the fact that $\Gamma^>_{\mathrm{top}}(\vert\vec{p}\vert)$ decays. In particular, we have a significant suppression for $\vert\vec{p}\vert/T\simeq O(10)$. 

\section{Conclusions and future outlooks}

In this talk we presented the computation of the spatial Fourier transform of the two-point correlation function of the topological charge density operator by means of a double extrapolation procedure. This has been done for the pure $\SU(3)$ gauge theory at $T\simeq1.24~T_c$. We observe a suppression of the correlator as the 3-momentum $\vert\vec{p}\vert/T$ is increased: this suggests a decreasing behaviour of $\Gamma^>_{\mathrm{top}}(\vec{p})$ for large momenta, according to the results of Ref.~\cite{Notari:2022ffe}. 

We are planning to perform the inversion of the correlator in order to establish the shape of the topological rate as a function of the axion 3-momentum. In the future we will extend the computation of the topological rate to the full QCD case, at the same temperatures studied in Ref.~\cite{Bonanno:2023thi}. This is necessary not only to put better constraints on the QCD axion phenomenology, but also to provide new insights on the structure of the QCD vacuum.

\section*{Acknowledgements}
We thank G. Villadoro for useful discussions. The work of C.~Bonanno is supported by the Spanish Research Agency (Agencia Estatal de Investigación) through the grant IFT Centro de Excelencia Severo Ochoa CEX2020-001007-S and, partially, by the grant PID2021-127526NB-I00, both of which are funded by MCIN/AEI/10.13039/501100011033. Numerical simulations have been performed on the \texttt{Leonardo} machine at Cineca, based on the agreement between INFN and Cineca, under project INF24\textunderscore npqcd. This work has also been partially supported by the project “Non-perturbative aspects of fundamental interactions, in the Standard Model and beyond” funded by MUR, Progetti di Ricerca di Rilevante Interesse Nazionale (PRIN), Bando 2022, grant 2022TJFCYB (CUP I53D23001440006). \sloppy

\providecommand{\href}[2]{#2}\begingroup\raggedright\endgroup

\end{document}